\documentstyle[twoside,fleqn,espcrc2,fps]{article}

\newcommand{\AmS}{{\protect\the\textfont2
  A\kern-.1667em\lower.5ex\hbox{M}\kern-.125emS}}

\def\lsi{\raise0.3ex\hbox{$<$\kern-0.75em\raise-1.1ex\hbox{$\sim$}}}
\def\gsi{\raise0.3ex\hbox{$>$\kern-0.75em\raise-1.1ex\hbox{$\sim$}}}
\newcommand{\lsim}{\mathop{\lsi}}

\newcommand{\R}{{\kern+.25em\sf{R}\kern-.78em\sf{I} 
  \kern+.78em\kern-.25em}}

\title{Simulating Simplified Versions of the IKKT Matrix Model}
\author{J.\ Ambj\o rn $^{{\rm a}}$, K.N. Anagnostopoulos $^{{\rm b}}$,
W. Bietenholz $^{{\rm c}}$, T. Hotta $^{{\rm d}}$ and J. Nishimura
\address{ Niels Bohr Institute, 
Blegdamsvej 17, DK-2100 Copenhagen \O, Denmark\\
$^{{\rm b}}$ Dept. of Physics, Univ.
of Crete, P.O. Box 2208, GR-71003 Heraklion, Greece \\
$^{{\rm c}}$ NORDITA, Blegdamsvej 17, DK-2100 Copenhagen \O, Denmark \\
$^{{\rm d}}$ Institute of Physics, Univ. of Tokyo,
Komaba, Meguro-ku, Tokyo 153-8902, Japan}
\thanks{Talk presented by W.B. at LATTICE 2000.}
}
\begin{document}

\begin{abstract}

We simulate a supersymmetric matrix model obtained from 
dimensional reduction of 4d SU($N$) super Yang-Mills theory 
(a 4d counter part of the IKKT model or IIB matrix model). 
The eigenvalue distribution determines the space structure.
The measurement of Wilson loop correlators reveals a universal 
large $N$ scaling. Eguchi-Kawai equivalence may hold
in a finite range of scale, which is also true for the bosonic 
case. We finally report on simulations of a low energy approximation 
of the 10d IKKT model, where we omit the phase of the
Pfaffian and look for evidence for a spontaneous Lorentz 
symmetry breaking.

\vspace*{-7mm}

\end{abstract}

\maketitle

\section{Motivation}

Several candidates for a constructive definition
of superstring theory have recently attracted attention.
Here we focus on the IKKT model (or IIB matrix model)
\cite{IKKT}, which is supposed to correspond to IIB 
superstring theory in the large $N$ limit.
%($N\times N$ being the size of the matrices).
That correspondence is supported by some analytical
arguments. This matrix model is formally
obtained from ordinary super YM gauge theory in the
zero volume limit (one point).

Here we study directly the large $N$ dynamics
of large $N$ reduced  matrix models. Some results were obtained
before for the ``bosonic case'' (where the fermions are dropped by 
hand), but we now want to address mainly the SUSY case.
In particular we simulate the 4d counterpart of the IKKT model 
\cite{pap1} --- we denote it as 4d IIB matrix model --- 
which corresponds again to the dimensional reduction 
of 4d super YM gauge theory.
%For an analytical approach to this model, see Ref.\ \cite{HKK}.
This model has also been studied analytically \cite{HKK} and
numerically in the framework of dynamical triangularization 
\cite{Bielefeld}.
Here we report on direct Monte Carlo simulations using the Hybrid R
algorithm \cite{HybR}. Conclusive results can be obtained 
because in the 4d version the fermion determinant is real positive 
--- in contrast to the 10d IKKT model, where simulations would
be plagued by a sign problem.
We got away with a computational effort of $O(N^{5})$
in the SUSY case, and of $O(N^{3})$ in the bosonic case.

\section{The 4d IIB matrix model}

The 4d IIB matrix model is given by 
%the partition function
\begin{eqnarray} \nonumber
{\bf Z} &=& \int dA \ e^{-S_{b}} 
\int d \bar \psi d \psi \ e^{-S_{f}} \\
S_{b} &=& -\frac{1}{4g^{2}} \, {\rm Tr} [A_{\mu},A_{\nu}]^{2}
\nonumber \\
S_{f} &=& -\frac{1}{g^{2}} \, {\rm Tr} ( \bar \psi_{\alpha} 
\Gamma^{\mu}_{\alpha \beta} [A_{\mu},\psi_{\beta}])
\end{eqnarray}
where $A_{\mu},\, \bar \psi_{\alpha},\, \psi_{\alpha}$
($\mu = 1 \dots 4,\, \alpha =1,2$)
are complex, traceless $N\times N$ matrices, and the
$A_{\mu}$ (only) are Hermitean.
We use $\vec \Gamma = i \vec \sigma, \, \Gamma_{4} = 1 \!\! 1$.
%,and the functional measure is taken over each matrix
%element, the traceless condition being implemented
%\cite{pap1}. 
%This model has a number of symmetries, which are 
%inherited from gauge theory before dimensional reduction:
In addition to SUSY and SO(4) invariance,
this model has a SU($N$) symmetry,
which is inherited from gauge invariance.
%(local and global coincides in $d=0$).

The first question about this model is if it is well-defined
as it stands.
Since the integration domain of $dA$ is non-compact, 
divergences are conceivable. However, our results, as well as
results on a number of special cases \cite{welldef,HNT,AIKKT} 
confirm consistently that this model {\em is} well-defined at 
any $N$; there is no need to impose a IR cutoff.
This implies that the only parameter $g$ is simply a
{\em scale parameter}, rather than a coupling constant.
It can be absorbed by introducing dimensionless quantities.
%\begin{equation} \label{dimless}
%X_{\mu} = A_{\mu} / \sqrt{g} \ ; \quad
%\Psi_{\alpha} = \psi_{\alpha}/g^{3/4} \, .
%\end{equation}
The challenge is, however, to tune $g$ as a function of $N$
so that the correlators are finite at $N \to \infty$,
%Various functions seem possible on theoretical grounds,
%but the question can be settled accurately based on
%our numeric results,
see Sections 4 and 5.

\section{The space structure}

In the IIB matrix model, the space coordinates arise dynamically
from the eigenvalues of the matrices $A_{\mu}$ \cite{IKKT}. In general
the latter cannot be diagonalized simultaneously, which implies
that we deal with a non-classical space.
We measure its uncertainty by
\vspace*{-1mm}
\begin{displaymath}
\Delta^{2} = \frac{1}{N} \Big[ {\rm Tr} (A_{\mu}^{2}) -
\, ^{max}_{U \in SU(N)} \sum_{i} 
\{ (UA_{\mu}U^{\dagger})_{ii} \}^{2} \Big] \ ,
\end{displaymath}
\vspace*{-1mm}
and the ``maximizing'' matrix $U$ is also used for introducing the
coordinates of $N$ points,
\vspace*{-1mm}
\begin{equation}
x_{i, \mu} = (U_{max} A_{\mu} U_{max}^{\dagger})_{ii} 
\qquad (i=1\dots N).
\end{equation}
\vspace*{-1mm}
What we are really interested in is their pairwise separation
$r (x_{i},x_{j}) = \vert x_{i}-x_{j} \vert$, and we show the
distribution $\rho (r)$ in Fig.\ 1.
We observe $\rho \approx 0$ at short distances 
($r/\sqrt{g} \lsim 1.5$), hence a UV cutoff is
generated {\em dynamically}. We also see that increasing $N$
favors larger values of $r$. To quantify this effect we measure
the ``extent of space''
%(note that $R^{2} = \int_{0}^{\infty}r^{2} \rho (r) \ dr$ diverges)
\footnote{Note that the quantity $R^{2} = \int_{0}^{\infty}
r^{2} \rho (r) \ dr$ diverges, 
since $\rho (r) \propto r^{-3}$ at large $r$ 
(see second Ref.\ in \cite{welldef}).}
\vspace*{-1mm}
\begin{equation}
R_{new} = \int_{0}^{\infty} r \rho(r) \ dr \ .
\end{equation}
\vspace*{-1mm}
Fig.\ 2 shows $R_{new}$ and $\Delta$ as functions of $N$
(at $g=1$). The inclusion of fermions enhances $R_{new}$
and suppresses $\Delta$, keeping their product approximately
constant. The lines show that both quantities follow the same
power law, $R_{new}, \ \Delta \propto N^{1/4}$, in SUSY 
and in the bosonic case. In SUSY this behavior is 
consistent with the branched polymer picture: there
one would relate the number of points as $N \sim
(R_{new}/\ell )^{d_{H}}$, where $\ell$ is some minimal bond,
which corresponds to the above UV cutoff.
The Hausdorff dimension $d_{H}=4$ then reveals consistency
with our result.
\begin{figure}[hbt]
%\vspace{-8mm}
\def\fpsangle{0}
\epsfxsize=65mm
\fpsbox{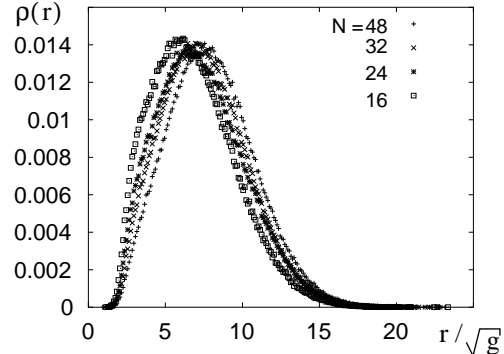}
\vspace{-12mm}
\caption{\it{The distribution of distances between 
space-points in the SUSY case at various $N$.}}
%\vspace{-9mm}
\end{figure}
\begin{figure}[hbt]
\vspace{-8mm}
\def\fpsangle{270}
\epsfxsize=40mm
\fpsbox{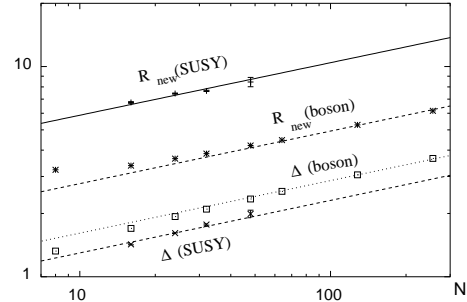}
\vspace{-11mm}
\caption{\it{The ``extent of space'' $R_{new}$ and the
space uncertainty $\Delta$ as functions of $N$ at $g=1$.}}
\vspace{-9mm}
\end{figure}

%Note that $\Delta$ remains finite at large $N$, which agrees
%with superstring theory (for a review, see Ref.\ \cite{Yoneya}).

\vspace*{-1mm}
\section{Polyakov and Wilson loops}
\vspace*{-1mm}

We define the Polyakov loop $P$
%as a Fourier transform of the eigenvalue distribution, 
and the Wilson loop $W$ --- which is conjectured to correspond to
the string creation operator --- as
\vspace*{-1mm}
\begin{eqnarray}
P(p) &=& \frac{1}{N} {\rm Tr} \Big( e^{i p A_{1}} \Big) , \\
W(p) &=& \frac{1}{N} {\rm Tr} \Big( e^{i p A_{1}}e^{i p A_{2}}e^{-i p A_{1}}
e^{-i p A_{2}} \Big) .
\vspace*{-1mm}
\end{eqnarray}
Of course the choice of the components of $A_{\mu}$ is irrelevant,
and the parameter $p \in \R$ can be considered as a ``momentum''.

Now $g(N)$ has to be tuned so that $\langle P \rangle ,\,
\langle W\rangle $ remain finite as $N\to \infty$. 
This is achieved by
\begin{equation} \label{g2N}
g \propto 1/\sqrt{N} \ ,
%\frac{1}{\sqrt{N}} \ ,
\end{equation}
which leads to a beautiful large $N$ scaling; Fig. 3 shows the invariance 
of $\langle P \rangle$ for $N=16 \dots 48$ in
SUSY. Also the bosonic case scales accurately \cite{pap1}.
\begin{figure}[hbt]
\vspace{-12mm}
\def\fpsangle{270}
\epsfxsize=40mm
\fpsbox{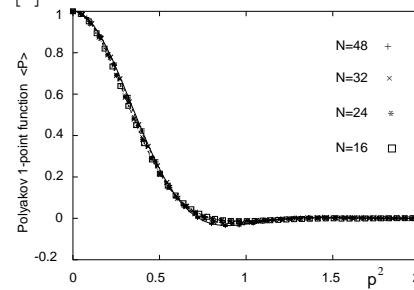}
\vspace{-10mm}
\caption{\it{The Polyakov function in the SUSY case
for various values of $N$ and $g^{2}N=const.$}}
\vspace{-8mm}
\end{figure}

The historic 2d Eguchi-Kawai model \cite{EK} had a 
$(Z\!\!\! Z_{N})^{2}$ symmetry, which implied
$\langle P(p \neq 0) \rangle = 0$, a property which was crucial
for the proof of the Eguchi-Kawai equivalence to gauge
theory. As we see, this property is not fulfilled here, but 
$\langle P(p) \rangle$ falls off rapidly, towards a regime where
the assumption of this proof holds approximately.

We proceed to a more explicit test of Eguchi-Kawai equivalence
by checking the area law for $\langle W(p) \rangle$.
Fig.\ 4 shows that the area law seems to hold in a finite
range of scale. Remarkably, the behavior is very
similar \cite{pap1} in the bosonic case. There we further investigated
the behavior at much larger $N$ \cite{prep}, 
and we observed that the power law regime does neither shrink 
to zero --- as it was generally expected --- nor extend to 
infinity --- a scenario which seems possible from Fig.\ 4.
At least in the bosonic case its range remains finite at large $N$.
Recently the same behavior was observed in a study of the 10d
bosonic case \cite{HE}.
\begin{figure}[hbt]
\vspace{-11mm}
\def\fpsangle{270}
\epsfxsize=40mm
\fpsbox{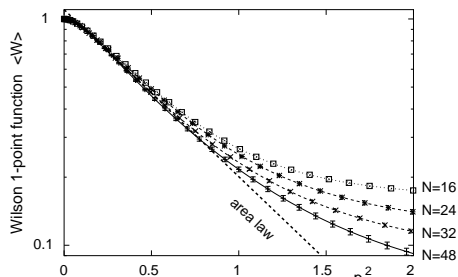}
\vspace{-11mm}
\caption{\it{The Wilson loop in the SUSY case
for various values of $N$ and $g^{2}N = const.$}}
\vspace{-12mm}
\end{figure}

\section{Multipoint functions}
\vspace*{-1mm}

We now consider connected multipoint functions
$\langle {\cal O}_{1} {\cal O}_{2} \dots {\cal O}_{n} \rangle_{con}$,
${\cal O}_{i}$ being a Polyakov or a Wilson loop.
We wonder if it is possible to renormalize all of those
multipoint functions simply by inserting ${\cal O}_{i}^{(ren)}
= Z {\cal O}_{i}$, so that a single factor $Z$ renders all
functions $\langle {\cal O}_{1}^{(ren)} {\cal O}_{2}^{(ren)} 
\dots {\cal O}_{n}^{(ren)} \rangle_{con}$ (simultaneously)
finite at large $N$.

It turns out that such a universal renormalization factor seems 
to exist in SUSY. We have to set again 
$g \propto 1/\sqrt{N}$, and then $Z \propto N$ provides
large $N$ scaling, as we observed for a set of 2, 3 and 4-point 
functions. Two examples are shown in Fig.\ 5. 
Our observation can be summarized by
the SUSY rule
\begin{displaymath}
\vspace*{-1mm}
\langle {\cal O} \rangle = O(1) \ , \quad
\langle {\cal O}_{1} \dots {\cal O}_{n} \rangle = O(N^{-n}) \quad
(n \geq 2).
\end{displaymath}
This implies that large $N$ factorization holds,
$\langle {\cal O}_{1} \dots {\cal O}_{n} \rangle = 
\langle {\cal O}_{1} \rangle \dots \langle {\cal O}_{n} \rangle
+ O(N^{-2})$, as in gauge theory, although
coupling expansions are not applicable here.
\begin{figure}[hbt]
\vspace{-9mm}
\def\fpsangle{270}
\epsfxsize=40mm
\fpsbox{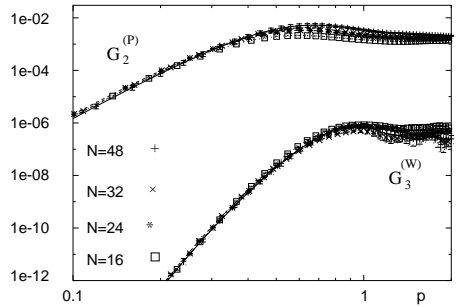}
%\vspace{-5mm}
%\def\fpsangle{270}
%\epsfxsize=40mm
%\fpsbox{wil3.susy.eps}
\vspace{-10mm}
\caption{\it{A Polyakov 2-point function
($G_{2}^{(P)}$ $= \langle [ {\rm Im} P(k)]^{2} \rangle$) 
and a Wilson 3-point function
($G_{3}^{(W)} = \langle [{\rm Im} W(k)]^{2} {\rm Re} W(k) \rangle
- \langle [{\rm Im} W(k)]^{2}\rangle \, 
\langle {\rm Re} W(k) \rangle$)
(both connected), with $g^{2}N=const.$ and renormalization factor
$Z\propto N$, which leads to large $N$ scaling in the SUSY case.}}
\vspace{-9mm}
\end{figure}

For the bosonic case, a $1/d$ expansion \cite{HNT} suggests
large $N$ factorization to hold as well, but it also predicts
$\langle {\cal O}_{1} \dots {\cal O}_{n} \rangle = O(N^{-2(n-1)}) \quad
(n \geq 2)$. This is confirmed numerically \cite{pap1}:
in particular the 3-point functions shown in Fig.\ 5 now
requires $Z^{3}\propto N^{4}$. Therefore
no universal renormalization factor $Z$ exists in the bosonic
case, which is an important qualitative difference from the
SUSY case.

\vspace*{-2mm}
\section{Simulations in ten dimensions}
\vspace*{-1mm}

We also performed simulations in $d=10$ \cite{pap2},
where we simplified the IKKT model as follows:

(1) We use a 1-loop approximation, which is expected to
capture the low energy dynamics. This amounts to an effective
action, keeping track of off-diagonal elements only to the
quadratic order, in the spirit of Ref.\ \cite{AIKKT}.

(2) We omit the phase of the Pfaffian by hand,
in order to avoid the sign problem.
Thus a Monte Carlo study becomes feasible.

The validity of (1) is supported by our results for $R_{new}$,
but (2) is certainly a drastic step. Still one could hope to observe basic
properties of the IKKT model at least qualitatively.
These simplifications allow for a simulation effort of only $O(N^{3})$.

Our main interest here is if the eigenvalue distribution
of the $A_{\mu}$ indicates a spontaneous symmetry
breaking (SSB) of SO(10) invariance,
%10d Lorentz invariance (or rotation invariance,
%since we are in Euclidean space), 
as it was suggested in the formulation of the IKKT model \cite{IKKT}.

To this end, we consider the moment of inertia
\vspace*{-3mm}
\begin{equation}
\vspace*{-3mm}
T_{\mu \nu} = \frac{2}{N(N-1)} \sum_{i>j} (x_{i \mu} - x_{j \mu})
(x_{i \nu} - x_{j \nu})
\vspace*{-3mm}
\end{equation}
($i=1\dots N$), and we measure its 10 eigenvalues.
A gap in this spectrum would indicate the SSB of Lorentz symmetry.
However, this cannot be observed, even though we raised $N$ up
to 512. On the contrary, we observe that the eigenvalue
distribution becomes more and more isotropic as $N$ increases,
see Fig.\ 6, and the same is true in $d=6$ \cite{pap2}.
We conclude that if SSB of Lorentz symmetry occurs in the
IKKT model, then it must be driven by the imaginary part of 
the action.
\begin{figure}[hbt]
\vspace{-8mm}
\def\fpsangle{270}
\epsfxsize=40mm
\fpsbox{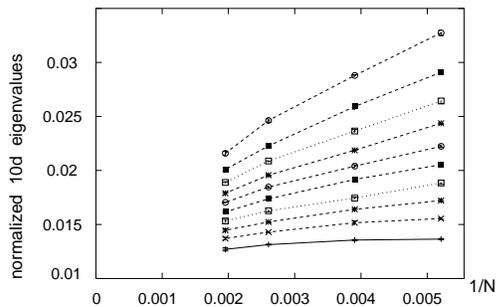}
\vspace{-10mm}
\caption{\it{The spectrum of the moment of inertia
(normalized by $\protect\sqrt{N}$) for $N=192,\, 256,\, 384
,\, 512$.}} 
%and 512.}}
\vspace{-12mm}
\end{figure}

\vspace*{-1mm}
\section{Conclusions}
\vspace*{-1mm}

We first simulated the 4d IIB matrix model, both, SUSY and bosonic.
We varied $N$ up to 48, which turned out to be sufficient to study
the large $N$ dynamics.

We confirmed that the model is well-defined as it stands,
hence $g$ is a pure scale parameter. 
%A UV cutoff is generated dynamically. 
The space coordinates arise from eigenvalues of the
bosonic matrices $A_{\mu}$. The extent of space
follows a power law. In SUSY this agrees with the branched polymer picture.
Fermions leave the power unchanged but reduce the space uncertainty
--- though it remains finite at large $N$.

The large $N$ scaling of Polyakov and Wilson loops and
their correlators requires $g \propto 1/\sqrt{N}$ in SUSY
and in the bosonic case, but the wave function renormalization
is qualitatively different: only in SUSY a universal renormalization
exists.

The area law for Wilson loops holds in a finite range of scale
for the SUSY and the bosonic case. The latter comes as a surprise,
and we checked up to rather large $N$ that this range remains
indeed finite. Hence Eguchi-Kawai equivalence to ordinary
gauge theory \cite{EK} may hold in some regime.

Finally we simulated a 10d low energy effective theory,
where the phase was dropped by hand. We could not observe any 
sign of a spontaneous breaking of Lorentz symmetry.\\

%As an outlook, one may still redo such simulation with twisted
%boundary conditions \cite{GAO}.\\

\vspace*{-3mm}
{\small W.B. would like to thank for the excellent organization of
LATTICE 2000, where he enjoyed discussions with 
A. Gonz\'{a}lez-Arroyo, U.-J. Wiese and J. Wosiek.}

\end{document}